\documentclass[5p]{elsarticle}

\pdfoutput=1

\usepackage[utf8]{inputenc}  
\usepackage[T1]{fontenc}       
\usepackage{amsmath}           
\usepackage{amssymb}           
\usepackage{caption}           
\usepackage[labelformat=simple]{subcaption} 
\usepackage{heppennames2}
\usepackage{localDefs}
\PassOptionsToPackage{hyphens}{url}
\RequirePackage[unicode=true,pdftex]{hyperref} 
\hypersetup{
    bookmarksnumbered=true,   
    breaklinks=true,          
    pdftoolbar=true,          
    pdfmenubar=true,          
    pdffitwindow=false,       
    pdfstartview={FitV},      
    pdfpagelayout=SinglePage, 
    pdfnewwindow=true,        
    colorlinks=true,          
    linkcolor=darkgreen,      
    citecolor=darkred,        
    filecolor=darkmagenta,    
    urlcolor=darkblue         
}
\RequirePackage[capitalise]{cleveref} 

\journal{Nuclear Instruments and Methods in Physics Research: Section A}

\begin{document}

\begin{frontmatter}


\title{Ultra-Fast Hadronic Calorimetry}

\author[fnal]{Dmitri Denisov}
\author[vinca]{Strahinja Lukić\corref{corrauth}}
\ead{slukic@vinca.rs}
\author[fnal]{Nikolai Mokhov}
\author[fnal]{Sergei Striganov}
\author[vinca]{Predrag Ujić}

\address[fnal]{Fermilab, Batavia IL, USA}
\address[vinca]{Vinča Institute, University of Belgrade, Serbia}
\cortext[corrauth]{Corresponding author}

\begin{abstract}
Calorimeters for particle physics experiments with integration time of a few ns will substantially improve the capability of the experiment to resolve event pileup and to reject backgrounds.
In this paper the time development of hadronic showers induced by 30 and 60~GeV positive pions and 120~GeV protons is studied using Monte Carlo simulation and beam tests with a prototype of a sampling steel-scintillator hadronic calorimeter. In the beam tests, scintillator signals induced by hadronic showers in steel are sampled with a period of 0.2~ns and precisely time-aligned in order to study the average signal waveform at various locations with respect to the beam particle impact. Simulations of the same setup are performed using the MARS15 code. Both simulation and test beam results suggest that energy deposition in steel calorimeters develop over a time shorter than 2~ns providing opportunity for ultra-fast calorimetry. Simulation results for an ``ideal'' calorimeter consisting exclusively of bulk tungsten or copper are presented to establish the lower limit of the signal integration window.
\end{abstract}


\end{frontmatter}

\section{Introduction}
  \label{sec:intro}
  
Detector systems at existing and future high energy collider experiments face increasing challenges related to event pileup and accelerator related backgrounds \cite{atlas_pII, CMS_pI}. An important tool for pileup and background rejection is the timing cut for the rejection of off-time signals. For example, the beam crossing interval option of 5~ns at the High Energy LHC, or FCC-hh would reduce pile-up by a factor of five with respect to the 25~ns option, provided that the detector integration time is shorter than the beam crossing interval. The relation between the energy resolution and pileup has also been approached in a simulation study within the CLIC $\text{e}^+\text{e}^-$ linear collider project \cite{CLIC_PhysDet_CDR}. 

The hadronic calorimetry is particularly challenging in this respect. Depending on the absorber material, hadronic showers may develop over several ten to several hundred ns. Part of the hadronic shower energy is spent on the nuclear binding energy in reactions releasing nucleons from the absorber nuclei. In the case of neutrons, the binding energy is recovered in neutron capture reactions, provided these occur within the volume of the calorimeter and within the signal integration time window. Otherwise the binding energy remains undetected. The energy carried by neutrinos produced in the shower is also invisible. The fluctuation of the total invisible fraction is one of the main components of the energy resolution of a calorimeter. At high event rates the late component of the hadronic shower energy deposition contributes to the background for subsequent events, complicating reconstruction.

The loss of neutral hadron energy is recovered using so called ``compensating'' absorber materials, like uranium. A consequence of this, however, is that the development time of the hadronic showers reaches several hundred ns \cite{Cald93}. On the other hand, hadronic calorimeters using steel or copper as absorber demonstrate lower levels of late energy deposition \cite{Vinz86}. The shower time structure of steel absorbers shows advantages over the more dense tungsten \cite{t3b}.

Our study seeks to understand the limits on the time window for the integration of the energy deposition of hadronic showers imposed by the shower development time in the calorimeter absorber material. To reach this goal we use Monte Carlo (MC) simulations and beam tests with a prototype of a steel-scintillator calorimeter. As the thickness of hadronic calorimeters typically exceeds 1~m, an important parameter of the shower development time is the time needed for the relativistic particle to traverse the calorimeter. In this article we study the shower development in terms of the \emph{local} time $\tred = t - t_0$, where $t_0$ is time when the particle incident on the calorimeter would have crossed the studied calorimeter layer if moving along a straight line at the speed of light. For a collider experiment, the signal integration window defined in local time implies that the readout system is capable of location dependent integration windows such that the signal integration in a given cell starts at the moment when a relativistic particle arrives at the cell from the beam interaction point along a straight line. Signal integration in local time has been used as the underlying assumption in the simulation studies for CLIC \cite{CLIC_PhysDet_CDR}. A short integration time window clearly requires a choice of the active calorimeter material with fast response. Such technologies exist, while a detailed discussion is beyond the scope of this paper.

A number of studies have previously addressed various aspects of the time development of hadronic showers \cite{Cald93, Vinz86, Aco91, Akc07}. Dedicated efforts have been made recently to measure the time structure of the hadronic showers and provide benchmarking input for the simulation tools \cite{t3b}. The focus of the present study is on the local time span for the full development of the shower at a given calorimeter depth, thus addressing the question of minimum required local integration time.

Simulations were performed using the MARS15 MC shower simulation code \cite{Mok95, Mok14}. The beam tests were performed at the Fermilab Test Beam Facility (FTBF) \cite{ftbf}. The accuracy of the measured shower time development is limited in our studies by properties of the scintillation counter used to measure energy deposition. Still, it will be shown that our setup is sufficiently sensitive to distinguish shower development times of the order of 1-2~ns. This provides key information about shower time development to verify the potential of the proposed method of signal integration.

\secref{sec:mars} describes the MARS15 software used for the simulation. 
\secref{sec:setup} describes the experimental setup, the data acquisition and the beam. Results for the test calorimeter are presented in \secref{sec:results}. \secref{sec:mars-ideal} presents the simulation of an ``ideal'' calorimeter consisting exclusively of tungsten or copper in order to establish the lower limit of the energy integration window. Conclusions are given in \secref{sec:conclusions}.

\section{MARS15 simulation code} 
  \label{sec:mars}
  
MARS15 \cite{Mok95, Mok14} is a general purpose, all-particle MC 
simulation code. It contains established theoretical models for strong, weak 
and electromagnetic interactions of hadrons, heavy ions, and leptons. Most 
processes in the code can be treated exclusively (analogously), inclusively (with the corresponding statistical weights) or in mixed mode. The exclusive approach is 
used in this study. In this case the hadron--nucleus interactions are modeled with the
LAQGSM event generator \cite{Mash08}.  The LAQGSM module in MARS15 is based on the 
quark-gluon string model above 10~GeV and intranuclear cascade, pre
equilibrium, and evaporation models at lower energies. The EGS5 code for 
electromagnetic shower simulation is used for energies from 1~keV to 
20~MeV, with a native MARS15 module used at higher energies. 

Ultimately all cascade particles transform energy to electrons through 
decays, inelastic, and elastic interactions with atomic electrons. Appropriate 
energy thresholds are applied to finish simulation in a
reasonable time (see below for details). If the energy of a particle becomes lower than the threshold, particle transport is not continued and the remaining kinetic 
energy is assumed to be deposited in the local medium without additional delay.
This is done for the majority of stable particles, nuclear recoils, heavy ions, and photons.
Negative particles can be captured by the atomic nuclei 
of the medium. They decay while in an atomic orbit, emitting photons in the event of orbital transitions or are absorbed by the nucleus, with delay of up to 80~ns 
for uranium and $2.2\,\mathrm{\upmu s}$ for hydrogen. Positive particles may annihilate 
(positron, antinucleon) or decay (pions, kaons, etc.). Neutrons are captured in $(n,\gamma)$ reactions and the photons from the 
capture reaction ultimately produce electrons. In all cases, 
electrons, protons, photons, neutrons, and 
neutrinos are present in the final phase of the cascade. 
The deposited energy in MARS15 consists of the ionization energy loss and the sub-threshold particle energies. As shown in detail below care is taken that the particle thresholds are sufficiently low to avoid bias in the results that might arise from the inclusion of non-ionization energy losses.

The electrons produced by the ionization of the medium are simulated 
by treating the ``soft'' and the ``hard'' collisions with atomic electrons
separately. The soft electrons are simulated by sampling their 
angular and energy distribution, while the relatively small 
number of hard collisions, producing the so-called ``delta-electrons'', is
treated by detailed simulation of the interaction kinematics \cite{Str05}. 
The time of the energy deposition at each simulation step is 
calculated as the time at the end of the step.

The results of the MARS15 simulation depend on the choice of threshold energies 
for different particles. This dependence can be studied 
by reducing the threshold energies. Default MARS15 threshold energies are: 
1~MeV for charged hadrons and muons, 
0.5~MeV for electrons and 0.1~MeV for photons and neutrons. We verified 
that calorimeter simulation results are stable when the
threshold energies are reduced by a factor of ten.


\section{Experimental setup}
  \label{sec:setup}
  \subsection{Test beam setup}
\label{sec:installation}

\begin{figure}[h]
\centering
   \includegraphics[width=\columnwidth]{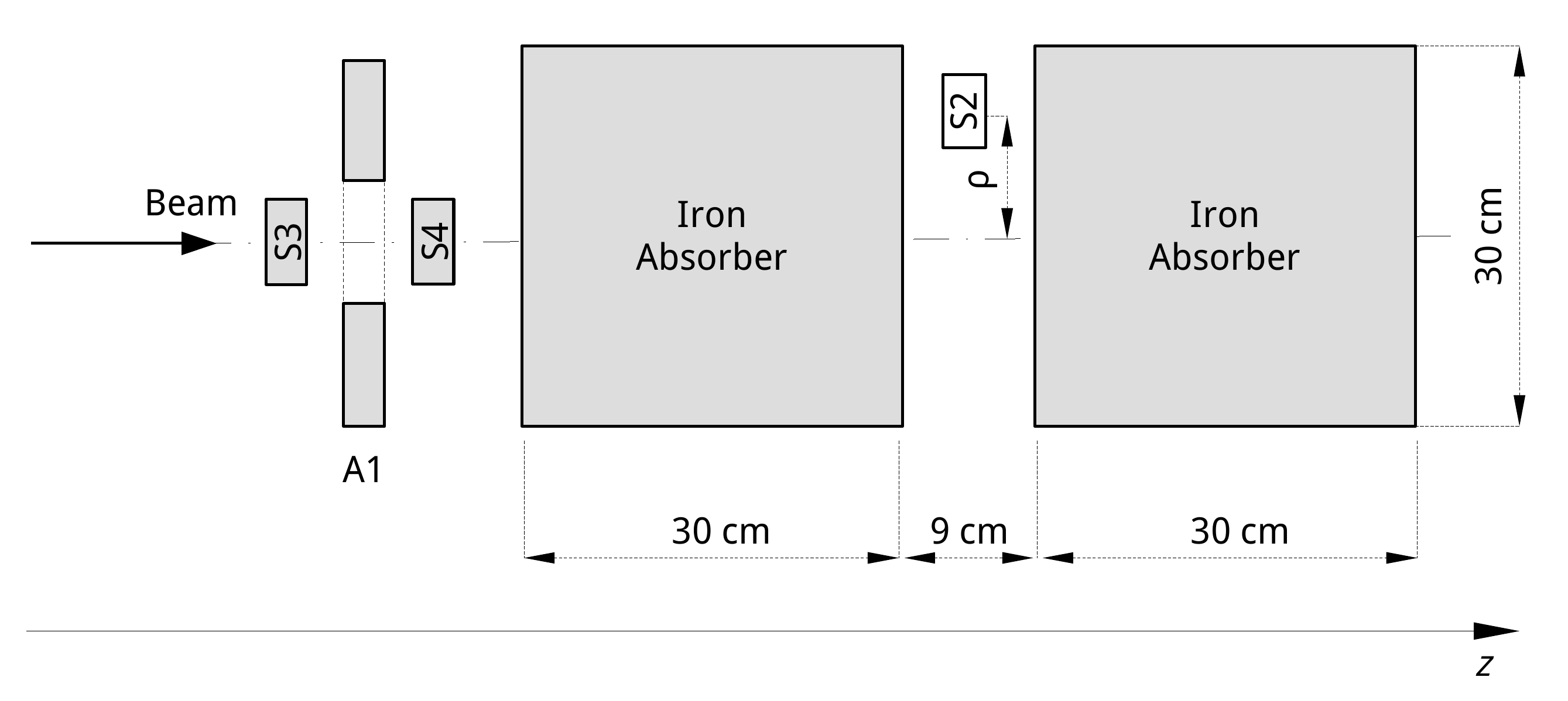}
   \caption{\label{fig:setup} Experimental setup (top view).}
\end{figure}

\figref{fig:setup} shows the top view of the experimental setup. Counter \stwo is 
placed between two iron absorber blocks to record local shower energy deposition. 
The cross section of both iron blocks is $30\times30\unit{cm}^2$, leaving 0.9 
interaction length, \lint, in the direction transverse to the beam between the 
beam impact point in the center and the closest edge of the absorber block. 
The total thickness of the absorber is 60~cm, corresponding to 3.6 $\lint$ in 
iron. In various runs,  the absorber thickness before and after the counter is 
subdivided into either $1.8+1.8\,\lint$ (as in \figref{fig:setup}) or $3.0+0.6\,
\lint$ (50~cm before and 10~cm after the \stwo counter). In the following, the 
configurations are referred to according to the thickness \emph{in front} of the 
counter. The total absorber thickness is always $3.6 \,\lint$. \tableref{tab:ez} 
shows the beam energies and absorber configurations used during the studies. 

The iron blocks are constructed of bricks and plates. Care was taken to avoid 
longitudinal joint slits in the construction along and near the beam axis. The 
density of the iron pieces was measured to be consistent with the density of 
steel, $7.7 \unit{g/cm}^3$, within the 2\% uncertainty of the measurement. 

The \stwo counter is assembled with \bicron scintillator material \cite{Bicron}, 
featuring fast response time and a FEU-115M photomultiplier tube \cite{Feu115M}, 
with good linearity and fast response. The FWHM of the signal induced by a MIP in 
the \stwo counter is 7.5~ns. The dimensions of the \stwo counter are $2.5\times15
\times1.25\unit{cm}^3$ in the horizontal direction perpendicular to the beam, 
vertical direction, and along the beam, respectively.

\begin{table}[h]
  \centering
  \caption{\label{tab:ez} Studied absorber thickness, beam energies, and particle types.}
  \begin{tabular}{ c c c c }
  \hline
  Front absorber thickness  &   30 GeV   &   60 GeV   &  120 GeV   \\
  \hline
  1.8 $\lint$   & \PGpp & \PGpp & \Pp \\
  3.0 $\lint$   &       &       & \Pp \\
  \hline
  \end{tabular}
\end{table}

Counters \sthr, \sfour and \textsf{A1} are positioned on the beam axis to trigger 
the data acquisition. The beam diameter at 10\% of the maximum is 1~cm. The 
dimensions of the counter \sthr are $2.5\times18\times1.25\unit{cm}^3$, of the 
counter \sfour $2.0\times6\times1.1\unit{cm}^3$ and of the counter \textsf{A1} 
$25.5\times25.5\times1.0\unit{cm}^3$. The dimensions are given in the horizontal 
direction perpendicular to the beam, vertical direction, and along the beam, 
respectively. Counter \textsf{A1} has a circular hole of 4~cm in diameter centered 
on the beam axis to veto upstream showers. The trigger logic is $\sthr\times\sfour
\times\overline{\textsf{A1}}$. Trigger signals are formed using NIM discriminator 
and coincidence modules.

\begin{figure}
\centering
   \includegraphics[width=\columnwidth]{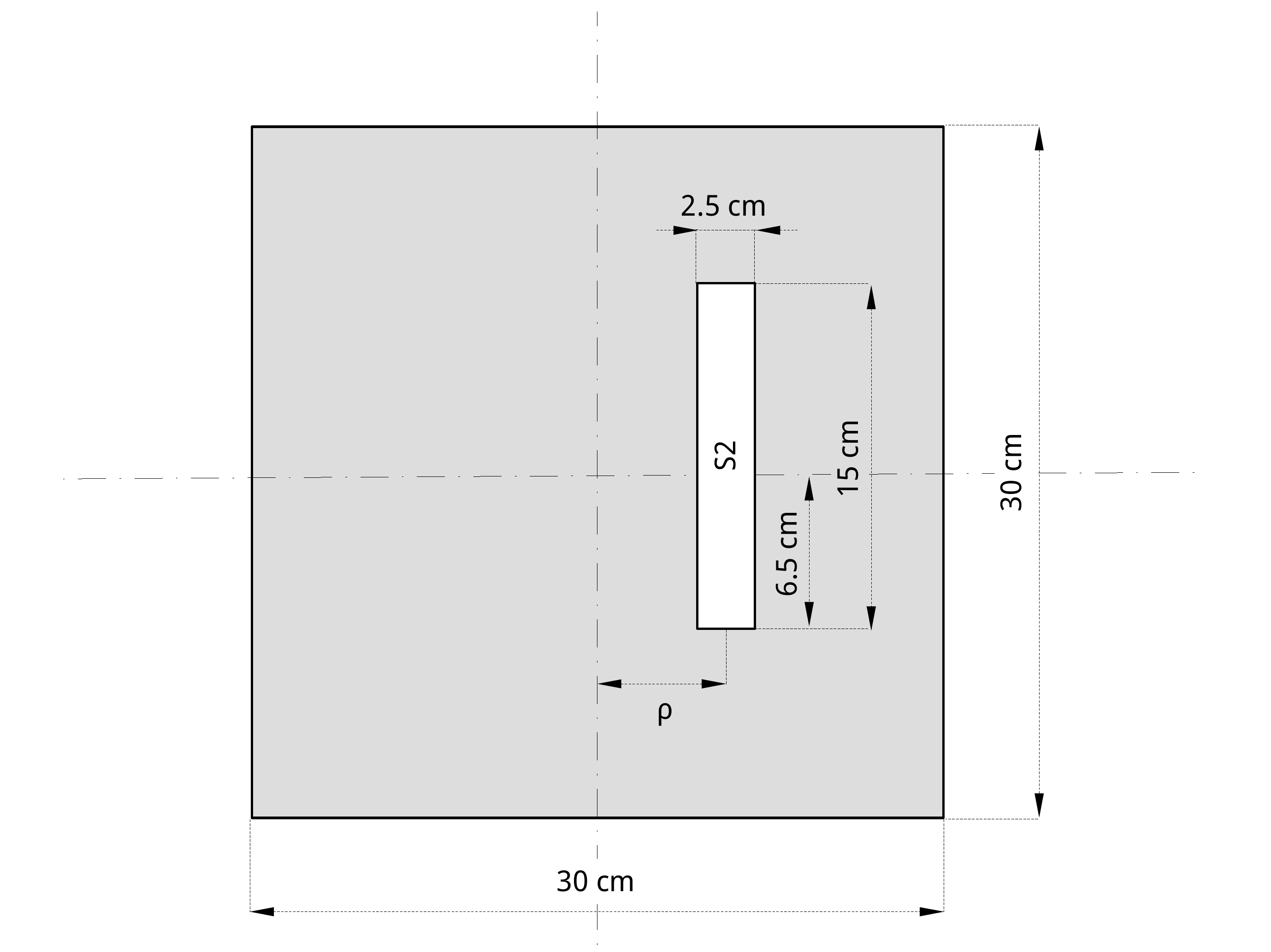}
   \caption{\label{fig:xy} The \stwo counter and the absorber in the transverse plane.}
\end{figure}

Fig.\ \ref{fig:xy} shows the cross-sectional layout of the setup including the 
relative position of the absorber and the counter. Three different distances from 
the beam axis, $\rho = 0$, 5 and 10~cm, are studied to scan the dependence of the 
energy deposition and of the time structure on the transverse distance from the 
shower core.  

The minimum ionizing particle (MIP) response of the \stwo counter was periodically 
recorded using the 120~GeV proton beam with the iron absorbers moved out of the 
beam.

\subsection{Data acquisition}
\label{sec:daq}

The signals of the scintillation counters are digitized using the DRS4 board, 
employing Switched Capacitor Arrays for high-speed signal digitization 
\cite{drs4}. The signals from counters \stwo, \sthr and \sfour are digitized at a 
rate of 5~GS/s. 
The reference time \tref for the passage of the beam particle is determined from 
the average time of  signals from the counters \sthr and \sfour, 
$\tref = (t_3 + t_4)/2$, using implemented offline digital constant-fraction 
discrimination. The precision of \tref was determined to be 110-120~ps, from the 
width of the distribution of the time difference $(t_3 - t_4)/2$.

The high voltage of the \stwo counter is adjusted so that the most probable 
amplitude for MIP signals is 7~mV. The low gain ensures good linearity of the PMT 
response and avoids saturation of the digitizer up to $\sim 70 \unit{MIP}$.

During the acquisition each of the four channels of the DRS4 board holds 200~ns 
of the signal waveform in a circular analogue buffer in the form of charges on an 
array of 1024 capacitors. When the trigger signal arrives, the chip digitizes the 
charges present in the buffer after a configurable delay. In our tests, the delay 
is configured in such a way that the digitized waveform contains $\sim 30\unit{ns}$ 
preceding the signal and 170~ns after the start of the signal. The first 21~ns of 
the waveform are used to calculate the baseline of the signal, the window from 27 
to 47~ns is used for the analysis of the deposited energy and the average signal 
shape, and the window from 47 to 180~ns is analyzed for the presence of 
additional pulses in the waveform. 

The MARS15 simulation uses a 20~ns integration window for energy deposition, as well.

  \subsection{Beam}
\label{sec:beam}

The beam at the FTBF is delivered once every 60~s in spills with a duration of 
4.2~s extracted from the Fermilab Main Injector. The primary beam consists of 
120~GeV protons. Several different beams are used in our studies:

\begin{enumerate}
  \item 120~GeV protons (primary beam).
  \item 60~GeV $\pi^{+}$ beam produced by the primary beam in a target, 450~m 
upstream from the test setup. A narrow band of particle momenta around 60~GeV is 
selected using dipole magnets and collimators. A small fraction of positrons is 
present in the beam, as well as muons from the pion decays.
  \item\label{lepi} 30~GeV $\pi^{+}$ produced by the primary beam in a target, 
160~m upstream from the test setup. A narrow band of particle momenta around 
30~GeV is selected using dipole magnets and collimators. Up to 10\% of positrons 
is present in the beam, as well as muons from the pion decays.
\end{enumerate}

The $\sthr\times\sfour\times\overline{\textsf{A1}}$ trigger rate during the tests 
is typically around 2.5~kHz. At this trigger rate, the acquisition rate of the 
DRS4 evaluation board is saturated at $\sim 500\unit{Hz}$, because of the limited 
data transfer rate via the USB bus. 

At the trigger rate of 2.5~kHz, including a 35\% $\overline{\textsf{A1}}$ veto 
rate, the probability that a second beam particle arrives during 150~ns after the 
trigger is about $6\times 10^{-4}$. The analysis of the recorded waveforms reveals 
that, indeed, a fraction of $6\times 10^{-4}$ of all waveforms contains a second 
pulse at a time delay corresponding to an integral multiple of the 19~ns bunch 
spacing of the Fermilab Main Injector \cite{ftbf-beam}.

\section{Results}
  \label{sec:results}
  \subsection{Deposited energy}
\label{sec:energy}

To establish the accuracy of the MARS15 simulation code, comparisons between the 
simulated and the measured distributions of the energy deposited in the counter 
\stwo are presented in this section. The energy deposited in the counter \stwo, 
$E_{\text{dep}}$, was measured by integrating the signal in the 20~ns window, as 
described in \secref{sec:daq}. Events in which a second pulse, induced by a beam 
particle from a different bunch, is present in the waveforms of either \sthr or 
\sfour are rejected. As the digitizer saturates for signals above $\sim70\unit{MIP}$, 
only events with $E_{\text{dep}}<70\unit{MIP}$ are analyzed. Both the experimental 
and the simulation results presented below are normalized to the unit sum of bin 
contents in the interval from 0 to 70~MIP. Geometrical details of the experimental setup, such as the spatial extent of the \mbox{\stwo} counter, were carefully modeled in the simulation.

\begin{figure}[h]
\centering
   \includegraphics[width=\columnwidth]{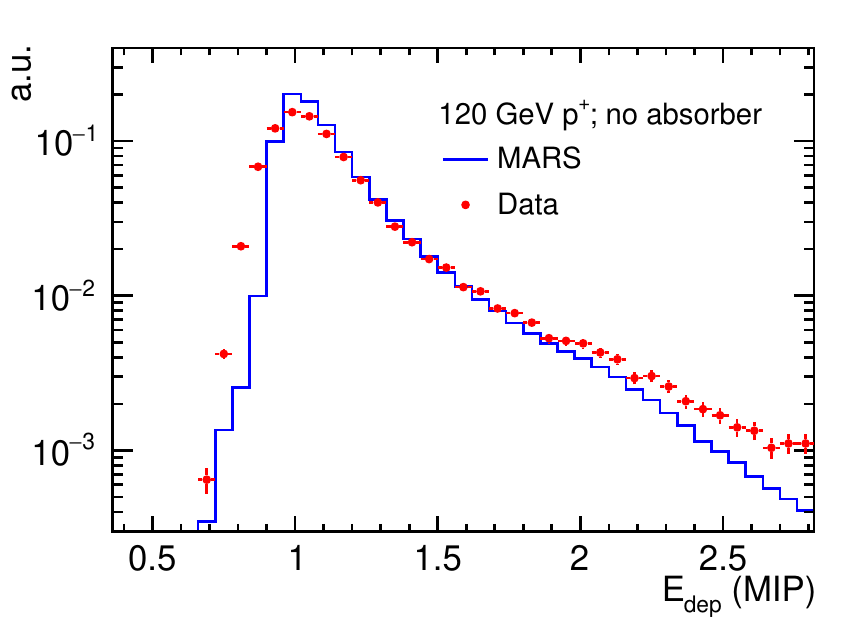}
   \caption{\label{fig:mip-units} Distribution of the deposited energy in the \stwo counter for a 120~GeV proton without absorbers (approximate MIP). The presented data uncertainties are statistical only. MARS15 statistical uncertainty per bin is at most two percent in the range $0.7\div2.8\unit{MIP}$.}
\end{figure}

\figref{fig:mip-units} shows the distribution of the deposited energy in the 
counter for 120~GeV protons without absorbers (approximate MIP). The experimental 
results and the MARS15 simulation are shown. The shapes of the distributions are 
in good agreement, taking into account that the experimental results contain the 
spread arising from the energy resolution of the counter, which is not covered by the simulation. The small excess in the 
experimental data around $E_{\text{dep}} = 2\unit{MIP}$ is due to double MIP 
depositions due to upstream showering and to occasional extraction of two protons 
in the same bunch from the Main Injector to the test beam facility. 

The most probable value (MPV) of the MIP deposited energy is used as the ``MIP unit'' for the comparison of experimental and simulation results. Dedicated MIP runs were repeated six times during the measurements to verify the stability of the MPV. In addition, MPV in runs with $\rho=0$ was also monitored. The MPV in experimental data exhibits run-to-run fluctuations with $\sim3\%$ relative standard deviation.

\begin{figure}
\centering
   \includegraphics[width=\columnwidth]{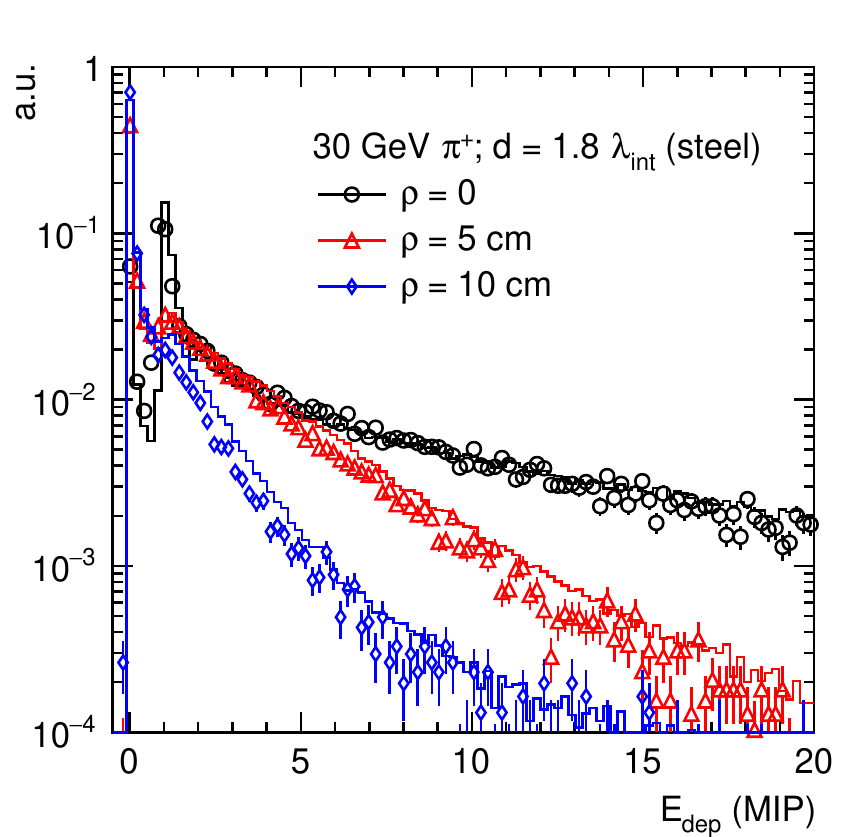}
   \caption{\label{fig:etot_1.8l_30G} Distribution of the deposited energies in the \stwo counter for 30~GeV pions after 1.8~\lint, at the three distances $\rho$ of the counter from the beam axis. Measurement results are represented by markers and simulation by the solid lines.}
\end{figure}

\begin{figure}
\centering
   \includegraphics[width=\columnwidth]{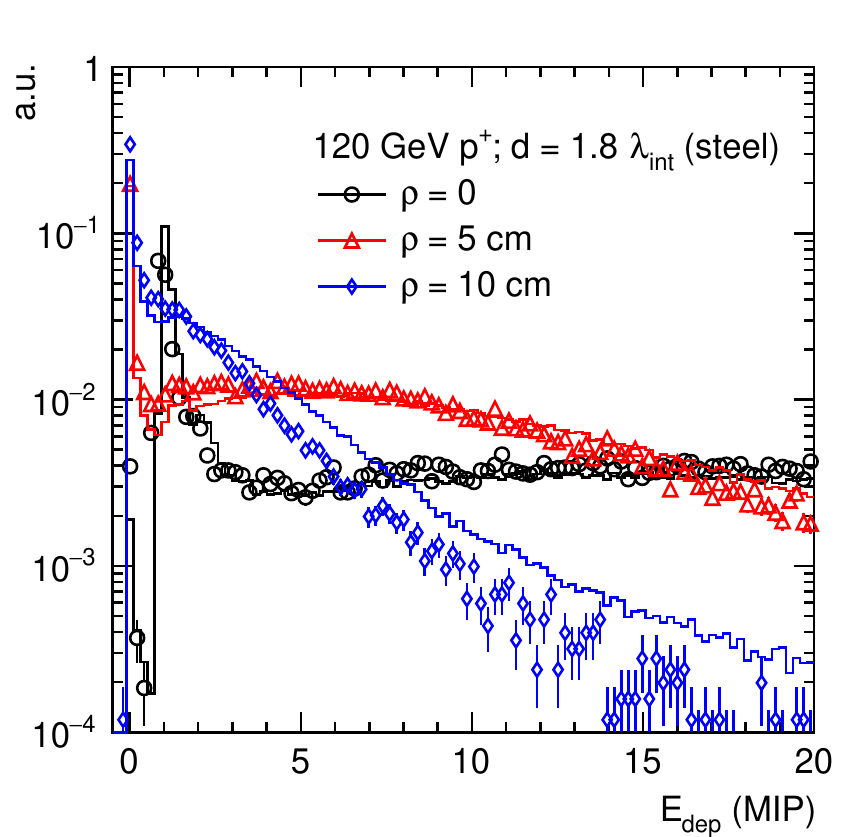}
   \caption{\label{fig:etot_1.8l_120G} Distribution of the deposited energies in the \stwo counter for 120~GeV protons after 1.8~\lint, at the three distances $\rho$ of the counter from the beam axis. Measurement results are represented by markers and simulation by the solid lines.}
\end{figure}

\begin{figure}[h]
\centering
   \includegraphics[width=\columnwidth]{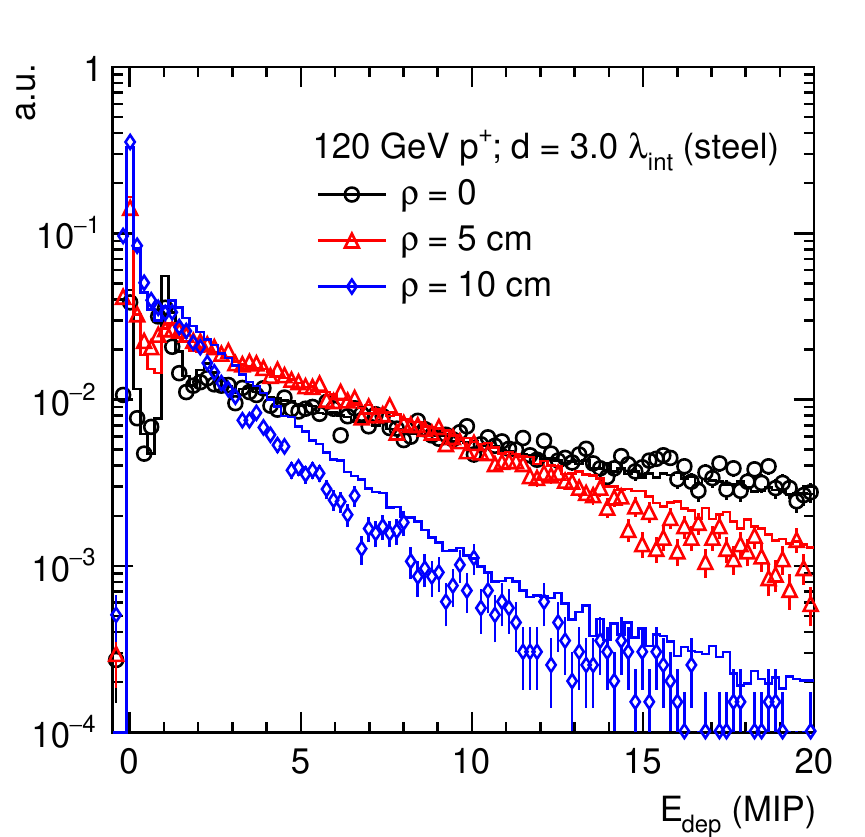}
   \caption{\label{fig:etot_3l_120G} Distribution of the deposited energies in the \stwo counter for 120~GeV protons after 3.0~\lint, at the three measured distances $\rho$ of the counter from the beam axis. Measurement results are represented by markers and simulation with the solid lines.}
\end{figure}

Figures \ref{fig:etot_1.8l_30G}-\ref{fig:etot_3l_120G} show the deposited energy 
distributions in the \stwo counter for various counter positions, absorber 
thicknesses, and beam energies. Figures \ref{fig:etot_1.8l_30G} and 
\ref{fig:etot_1.8l_120G} are for a 30~cm (1.8 \lint) steel absorber while 
\figref{fig:etot_3l_120G} is for a 50~cm (3 \lint) one. Each figure compares the 
data (points) and simulation results (histograms) at three measured distances 
$\rho$ of the counter from the beam axis: 0 (on axis), 5, and 10~cm. Good overall 
agreement is observed between the measurements and simulation, while at higher 
beam energies the simulation predicts slightly broader transverse profile of the 
shower than measured. As the fraction of energy deposited at large distance from 
the shower core is relatively small, the observed differences do not affect the 
main results of our studies.

One may notice a narrow peak at zero energy in all measured and simulated distributions, besides the broader peak at 1~MIP visible in the distributions for $\rho=0$ and $\rho=5\unit{cm}$. The zero peak for $\rho=0$ may be attributed to showers that start by a production of an energetic neutron, accompanied by the scattering of charged particles at angles sufficient to miss the counter. For higher values of $\rho$, the probability for the shower to deposit very low or zero energy in the S2 counter rises for geometrical reasons. The counter S2 may in that case be hit by soft particles at the fringe of the shower, or entirely missed. As opposed to the case of the single energetic neutron at $\rho=0$, this case involves a continuum of possible energy deposits close to zero. This explains why the zero energy peak for $\rho>0$ is broader than for $\rho=0$.

  \subsection{Shower time structure}
\label{sec:time}

To study the time structure of the hadronic shower energy deposition, the signal 
waveforms recorded in the \stwo counter are precisely time-aligned and averaged. 
The average signal waveform is calculated from several ten thousands of sampled 
signals for each run configuration characterized by the beam particle type and 
energy, and by the location of the \stwo counter. The time-alignment is performed 
by shifting the time of all digitized samples by $C - \tref$, where \tref is the 
reference time calculated as the average time of the digitized signals of \sthr 
and \sfour (See \secref{sec:daq}), and $C$ is a constant that defines the 
absolute time offset for all measurements. The constant $C$ is arbitrary in the 
sense that the choice of the physical time represented by $t=0$ is arbitrary. It 
is only important that $C$ is the same across all measurements. For 
the present analysis, we selected $C$ so that $t=0$ corresponds to the time when the average signal waveform begins deviating from the baseline.

In the MIP runs the \stwo counter is kept on the beam axis, at the same 
position along the beam as in the corresponding runs with the absorber. Since the 
MIP energy deposition is practically instantaneous, the average MIP signal 
waveform represents the response function of the \stwo counter in the time 
domain. 
When calculating the average signal waveforms for the MIP runs, signals with the 
measured energy deposition below 0.5~MIP are excluded. The FWHM of the average 
MIP signal is 7.5~ns and is limited by the response of the scintillator and the 
PMT.

\begin{figure}[h]
\centering
   \begin{subfigure}{\columnwidth}
   \includegraphics[width=\textwidth]{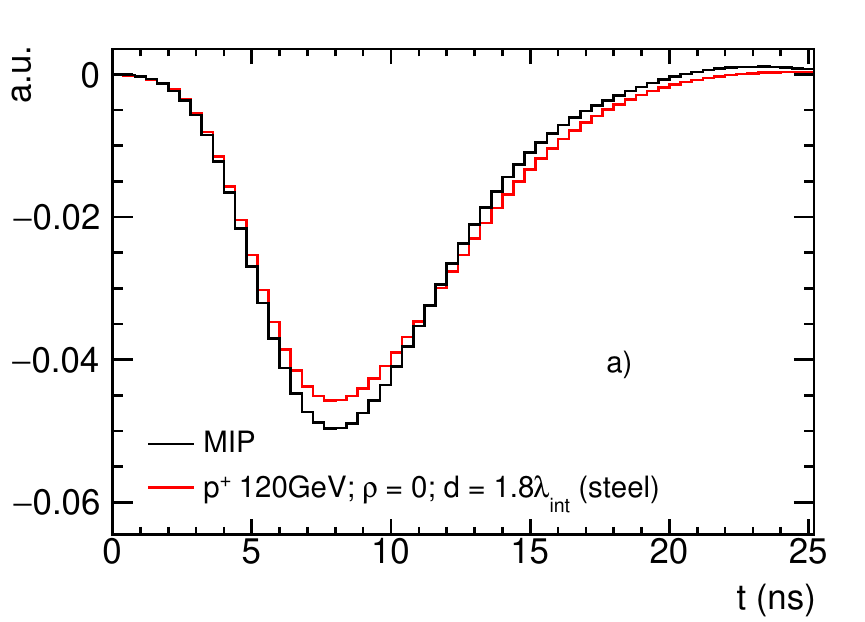}
   \end{subfigure}
   \begin{subfigure}{\columnwidth}
   \includegraphics[width=\textwidth]{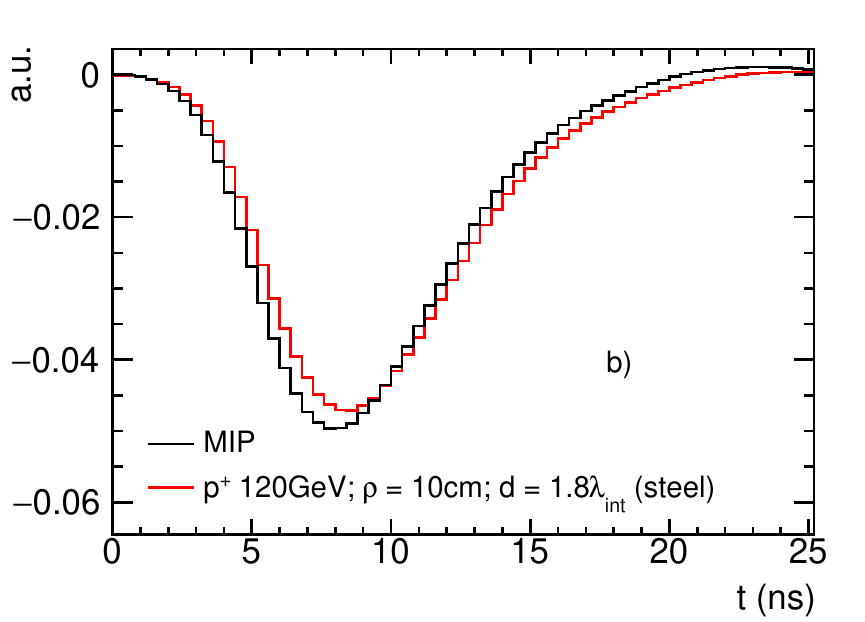}
   \end{subfigure}
   \caption{\label{fig:pulse_1l8_120G} Comparison of the average signal shapes from showers generated by 120~GeV protons after 1.8~\lint, with the average signal shape from MIPs: a) $\rho = 0$, b) $\rho = 10\unit{cm}$.}
\end{figure}

\begin{figure}[h]
\centering
   \begin{subfigure}{\columnwidth}
   \includegraphics[width=\textwidth]{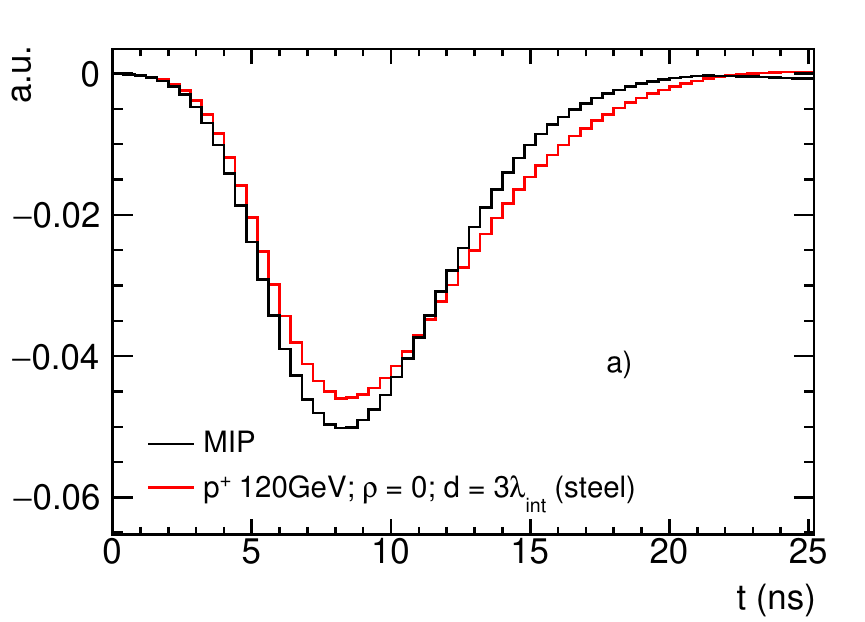}
   \end{subfigure}
   \begin{subfigure}{\columnwidth}
   \includegraphics[width=\textwidth]{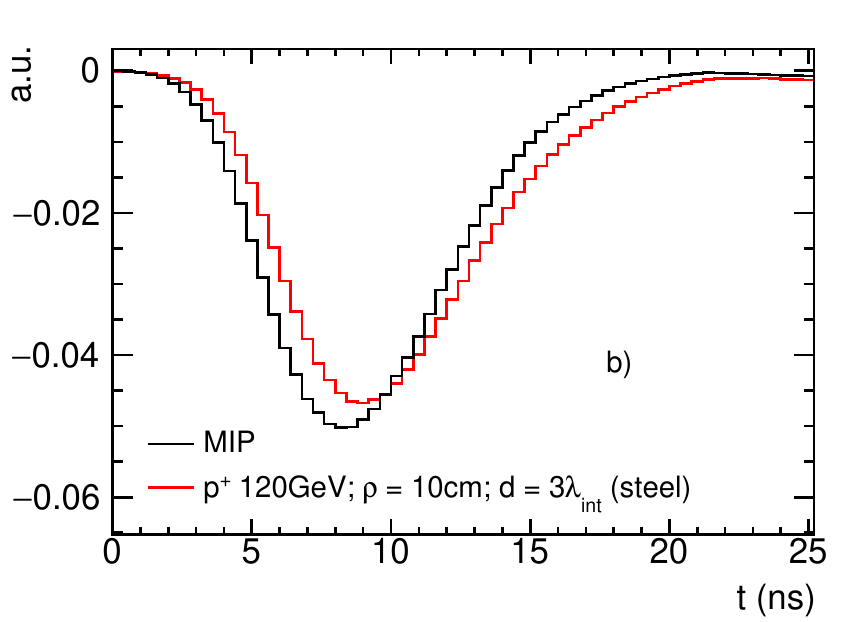}
   \end{subfigure}
   \caption{\label{fig:pulse_3l_120G} Comparison of the average signal shapes from showers generated by 120~GeV protons after 3~\lint, with the average signal shape from MIPs: a) $\rho = 0$, b) $\rho = 10\unit{cm}$.}
\end{figure}

Figures \ref{fig:pulse_1l8_120G} and \ref{fig:pulse_3l_120G} show the comparison 
of the average time-aligned signals induced by hadronic showers to the average 
signals induced by MIPs for several positions of the \stwo counter and the beam 
energy 120~GeV. Plot (a) on both figures shows the comparison with the counter on 
the beam axis and plot (b) for the counter 10~cm away from the axis. The hadronic and the MIP pulses are normalized to the same integral in the plots. The difference between the hadronic and the MIP pulses, while modest, increases with the absorber thickness.

The signal waveforms from the hadronic showers are slightly wider than those from the MIP events. This is due to the fact that the hadronic showers develop over a certain finite time, while the MIP energy deposition is practically instantaneous, with time duration of $d/c \approx 40\unit{ps}$, where $d$ is the thickness of the sensitive volume of the counter, and $c$ is the speed of light. However, the difference in the signal shape between a MIP and a hadronic shower is small. This indicates fast hadronic shower energy deposition. The MIP signals used for comparison are recorded at the moment of the MIP passage through the same calorimeter depth, implying that we are studying the development of the energy deposit in \emph{local time}.

The small delay of the hadronic signal registered at $\rho=10\unit{cm}$ from the beam axis with respect to the MIP signal is due to the time needed for the shower to develop at the position of the counter.

\subsubsection{Deconvolution of the time structure of the hadronic shower}

In order to estimate the shower development time, a deconvolution of the MIP response function from the averaged hadronic shower waveform is performed. This procedure has relatively large uncertainties due to the following:
\begin{itemize}
    \item The local shower development time is much shorter than the response function of the \stwo counter. This implies that the imperfections in the registered waveforms have strong impact on the results.
    \item The tail of the \stwo counter signal has low-amplitude ringing due to reflections in the signal path. The reflections are time dependent and are affected by various factors such as cabling connections. 
    \item At large amplitudes the nonlinearity of the \stwo counter response distorts the hadronic shower signals.
\end{itemize}
Nevertheless, the deconvolution procedure provides important information about characteristic times of a hadronic shower energy deposition.

\begin{figure}[h]
\centering
  \includegraphics[width=\columnwidth]{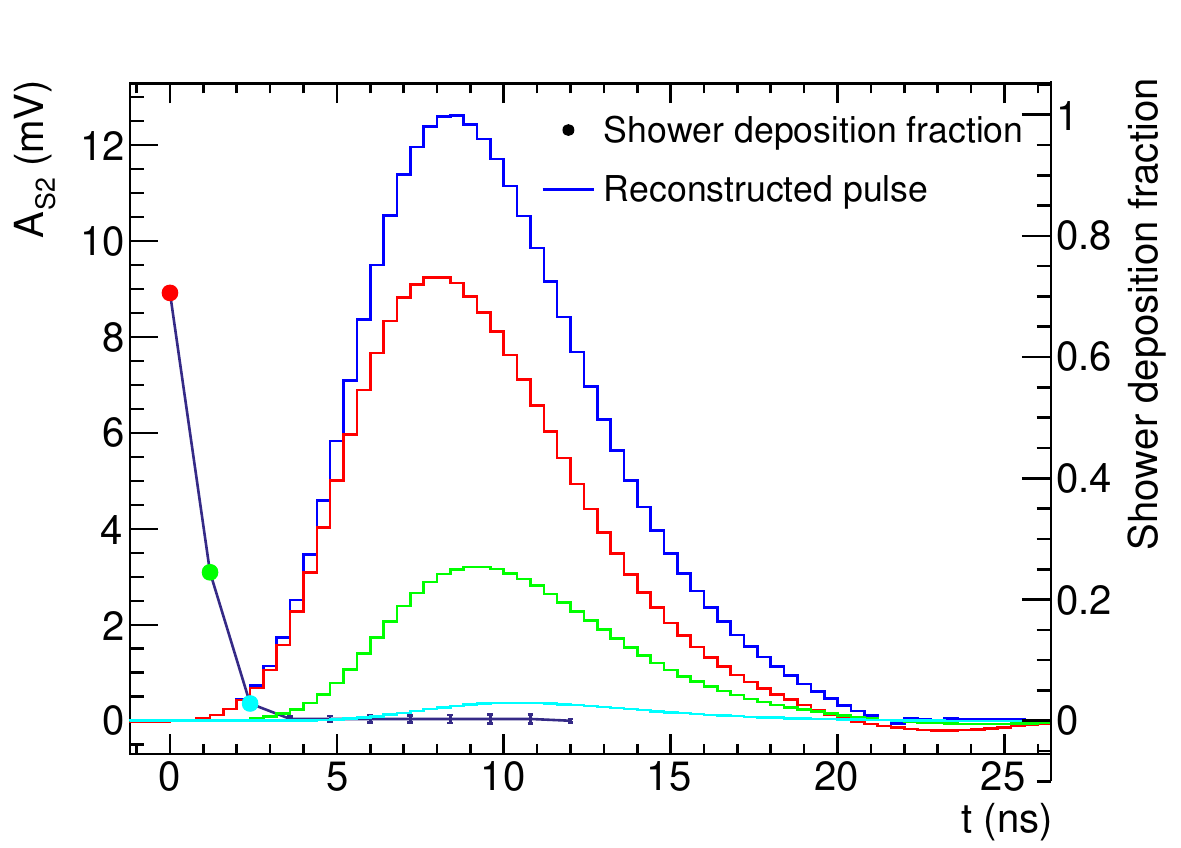}
  \caption{\label{fig:scheme} Discrete convolution of the shower energy deposition (dots) and MIP response (colored lines except blue) resulting in the reconstructed hadronic signal (blue line). $A_\stwo$ is the amplitude of the \stwo counter signal.}
\end{figure}

The deconvolution is performed by modeling the \stwo signal induced by a 
hadronic shower as a discrete convolution of the local time structure of the 
shower and the MIP response of the \stwo counter. The time structure of the 
shower is represented as a series of fractional energy depositions of the shower 
on a discrete time grid. The convolution model is  fitted to the averaged 
measured hadronic signal to extract the shower time structure. The concept is 
illustrated in \figref{fig:scheme}. The dots represent the fractions of the 
shower energy deposition on the discrete time grid. The scale for the deposition 
fraction is given on the right vertical axis. Various lines, except the blue, 
represent the MIP response of the \stwo counter scaled by the corresponding 
shower energy deposition fraction and delayed by the corresponding energy 
deposition time. The blue line represents the reconstructed signal of the 
hadronic shower obtained by summing the scaled and delayed MIP response signals. 
The period of the time grid is fixed to 1.2~ns. The only assumption made about 
the shower time structure is that the fractions decrease monotonically with time.

\begin{figure}
\centering
  \includegraphics[width=\columnwidth]{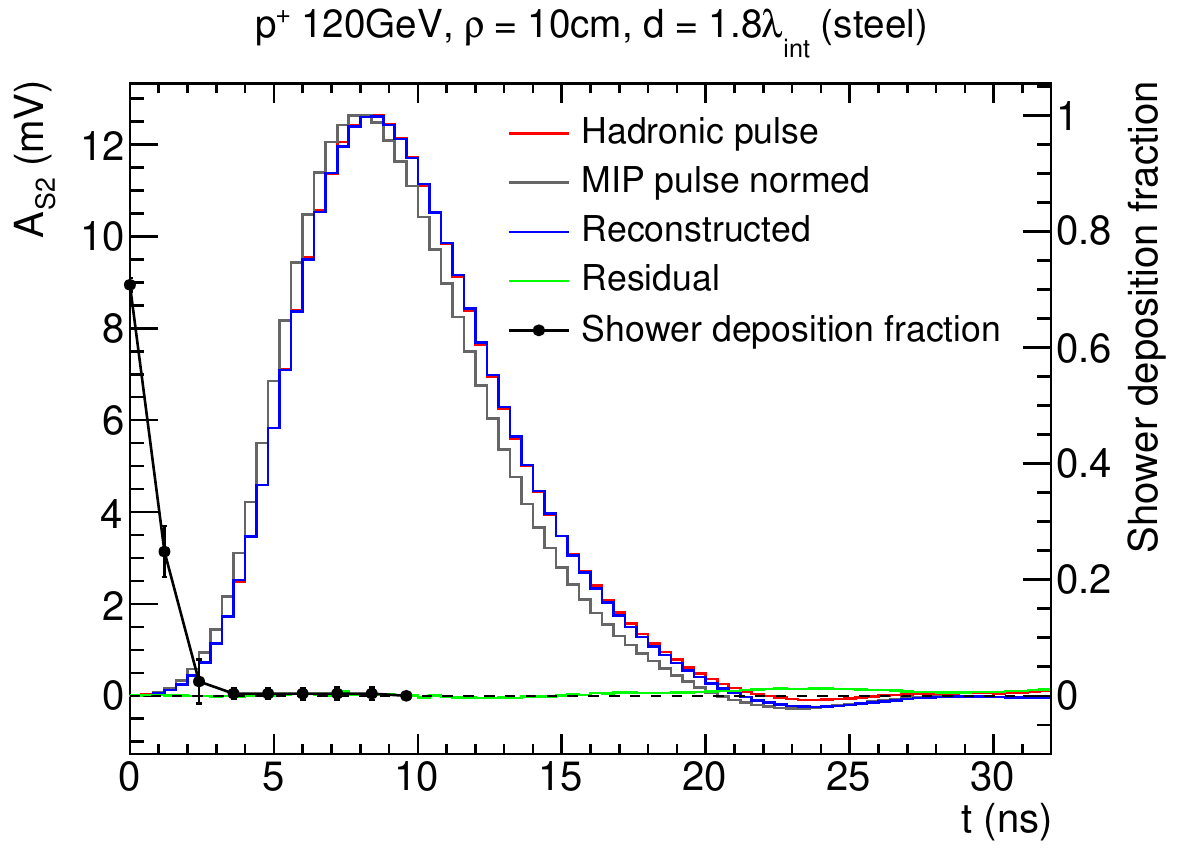}
  \caption{\label{fig:fit_1l8_120G} Fit of the shower energy deposition time distribution for the 120~GeV proton run behind 1.8~\lint and with the \stwo counter 10~cm off the beam axis. $A_\stwo$ is the amplitude of the \stwo counter signal.}
\end{figure}

\figref{fig:fit_1l8_120G} shows the fit of the model to the average signal shape for the 120~GeV protons behind 1.8~\lint and with the \stwo counter at $\rho=10\unit{cm}$ from the beam axis. The MIP response is represented by the gray line, the measured hadron shower signal by the red line, and the fitted deposition time distribution is shown as black dots. The MIP response has been scaled to the same peak amplitude as the hadronic signal. A reconstruction of the hadronic signal by convolution of the shower energy deposition time function and the MIP response is shown by the blue line. The residual difference between the reconstructed and the measured hadronic shower signal is shown as the green line.

We use the time needed to deposit 80\% of the total deposited energy in the \stwo counter, $t_{80\%}$, to compare experimental results with MARS15 simulation. In the experimental results $t_{80\%}$ is estimated from the fitted time distributions of hadronic shower energy depositions by linear interpolation between the points. The 80\% threshold is not too sensitive to the imperfections in the tail of the counter signal, while still providing the time needed for the deposition of a substantial fraction of the shower energy. 

As the present study is sensitive on the variation of the response of the counter due to, e.g.\ environmental conditions, we made an estimate of the effect of the counter response variation on $t_{80\%}$. 
As the deconvolution procedure involves both the average MIP signal waveform and the average hadronic signal waveform, the effect of the variation of the counter response 
on both has to be taken into account. 
The effect on $t_{80\%}$ due to the variation of the MIP signal waveform is estimated by repeating the fit procedure with various MIP runs taken at different times during the measurements. 
The standard deviation of $t_{80\%}$ in the repeated fit procedure is taken as the contribution of the MIP signal waveform variation to the uncertainty of $t_{80\%}$. 
With the available experimental data it is not possible to perform the same kind of analysis of the effect of the response variation on the hadronic signal waveforms. We assume that the contributions to the $t_{80\%}$ uncertainty from the variations on the hadronic and the MIP waveforms are equal and uncorrelated. Thus the total estimate for the uncertainty from the variation of the counter response is obtained by multiplying the MIP contribution by $\sqrt{2}$.

Because of the small number of MIP runs available, the uncertainty due to the response variation is not estimated for each \stwo counter position separately, but the average variance of the $t_{80\%}$ over all the measurements is used to make a single global estimate of this source of uncertainty at 0.3~ns.

Finally, the uncertainty arising from the finite bin width used for the deconvolution of the hadronic energy deposition time distribution is estimated as the RMS of a box function with the width equal to the bin width. This contribution equals 0.35~ns for all measurements.

The maximum analyzed signal amplitude is 70~MIP, which corresponds to a maximum signal amplitude of 500~mV at the phototube output. It is conceivable that the phototube saturation has an influence on the averaged hadronic signal shape. Such an effect should be most pronounced for high incident energy and at $\rho=0$. The small variation of measured $t_{80\%}$ for different incident energies at $\rho=0$ indicates that the saturation effect is small in comparison to other sources of measurement uncertainty.

\begin{table}
\caption{\label{tab:t80}
Time needed to deposit 80\% of the total energy in the \stwo counter.}
    \begin{tabular}{ c | c | c | c | c }
    \multicolumn{2}{c}{ } & \multicolumn{3}{|c}{$t_{80\%}$ (ns)} \\
    \hline
    Beam  &  s (\lint) & $\rho=0$ & $\rho=5\unit{cm}$ & $\rho=10\unit{cm}$ \\
    \hline
    \PGpp 30~GeV & 1.8 & 0.4 & 0.5 & 0.6 \\
    \PGpp 60~GeV & 1.8 & 0.6 & 1.2 & 1.7 \\
    \Pp  120~GeV & 1.8 & 0.5 & 0.4 & 1.1 \\
    \Pp  120~GeV & 3.0 & 0.6 & 1.6 & 1.2 \\
    \hline
    \end{tabular}
\end{table}

\begin{figure}
\centering
   \includegraphics[width=\columnwidth]{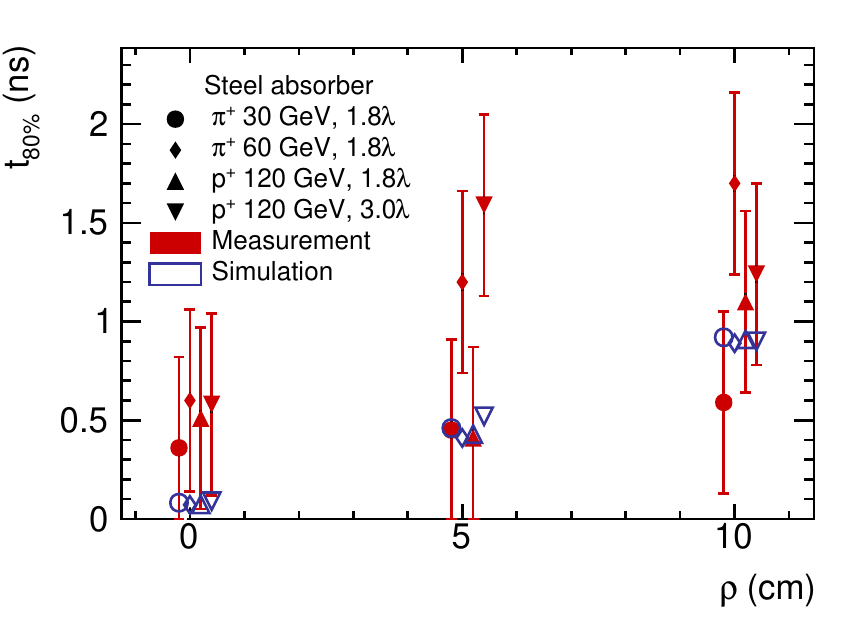}
   \caption{\label{fig:t80} $t_{80\%}$ for various beam energies 
as a function of the transverse position of the counter. 
Measurement and simulation results are shown. Small horizontal shifts are 
applied to the points for better visibility. The meaning of the symbols is as 
follows: circle -- $\pi^+$, 30~GeV, 1.8~\lint, diamond -- $\pi^+$, 60~GeV, 
1.8~\lint, triangle-up -- $p^+$, 120~GeV, 1.8~\lint,  triangle-down -- $p^+$, 
120~GeV, 3.0~\lint; full red symbols -- data, open blue symbols -- MARS15.}
\end{figure}

\begin{figure}
\centering
   \includegraphics[width=\columnwidth]{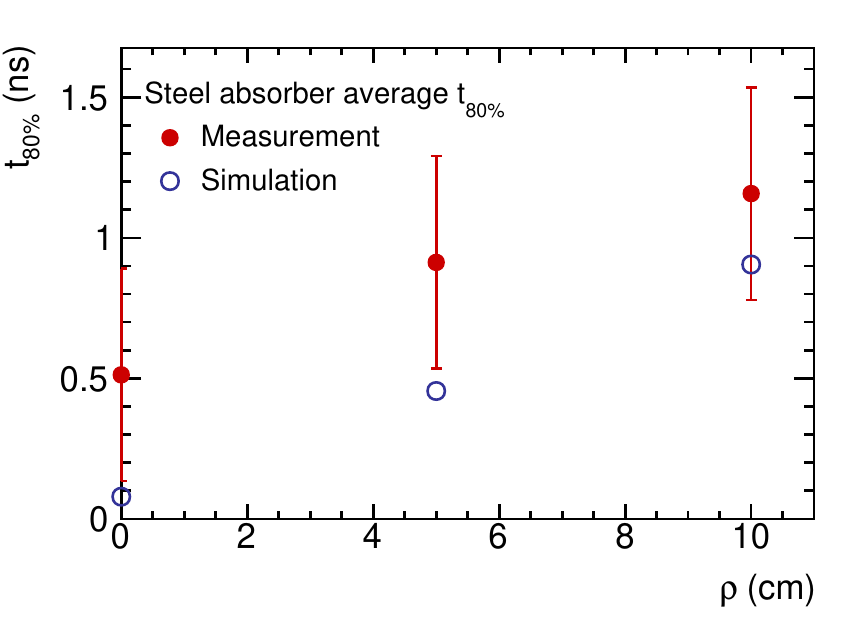}
   \caption{\label{fig:t80-avg} $t_{80\%}$ for various beam energies and positions of the counter. Results are averaged over the various particle types, beam energies, and different depths of the absorber.}
\end{figure}

\tableref{tab:t80} shows the measurements results for $t_{80\%}$ for all studied points.
\figref{fig:t80} compares the experimental and the simulation results for 
$t_{80\%}$ for various beam energies and positions of the counter. 
\figref{fig:t80-avg} compares the averaged simulation and experimental results 
for $t_{80\%}$ for various beam energies as a function of the transverse position 
of the counter. The uncertainties of the averaged experimental results are calculated under the assumption that the uncertainty arising from the finite bin width used in the deconvolution is common for all measurements, while the uncertainty arising from the variation of the counter response is fully uncorrelated between the measurements.

Figures \ref{fig:t80} and \ref{fig:t80-avg} indicate that MARS15 simulation of 
hadronic shower time development is in good agreement with the experiment and 
that the development of hadronic showers in iron is fast, with typical times of a 
few ns. 
The main finding is that 80\% of hadronic shower depositions in iron are deposited faster than $\sim 2\unit{ns}$.

\section{Energy resolution of a hadronic calorimeter with ultra-short integration time}
  \label{sec:mars-ideal}
  The presented comparisons of measured and simulated energy and time distributions 
of hadronic shower energy deposition give sufficient grounds to study energy 
deposition in the calorimeters based on the MARS15 simulation. 
In order to assess the ultimate limit of the resolution of a hadronic calorimeter 
with ultra-short local integration time, 
we perform simulations of ``ideal'' calorimeters consisting entirely of bulk tungsten and copper using MARS15 simulation code. The size of the simulated ``ideal'' calorimeters is selected to ensure that leakage effects are negligible. The resulting energy resolution does not contain the sampling term. The resolution is thus almost exclusively determined by the fluctuations of the hadronic fraction of the shower energy and the fluctuations of the time distribution of the energy depositions, which are the most important factor determining the effect of the short integration times on the energy resolution of the calorimeters. 
It is assumed that the deposited energy is read out instantaneously at the moment of the deposition.

\begin{figure}
\centering
    \includegraphics[width=\columnwidth]{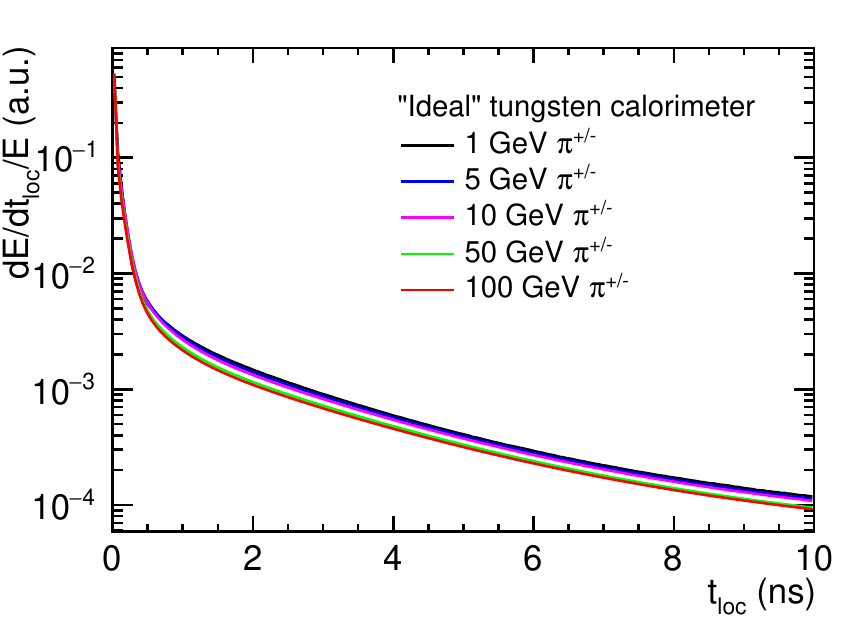}
    \caption{\label{fig:tref}Simulated fraction of the deposited energy in bulk tungsten for various incident pion energies as a function of \tred.}
\end{figure}

In the first simulation, the ``calorimeter'' is a tungsten cylinder with a length of 120~cm, corresponding to $\sim12\,\lint$, and a radius of 60~cm, corresponding to $\sim6\,\lint$. 
The showers are initiated by charged pions with various
incident energies. Simulated fractional distributions of the shower energy depositions as a function 
of \tred are presented in \figref{fig:tref}. These results indicate that the time 
dependence of the deposited energy is practically the same for all energies studied 
and that the hadronic energy deposition is very fast. 

\begin{figure}
\centering
    \includegraphics[width=\columnwidth]{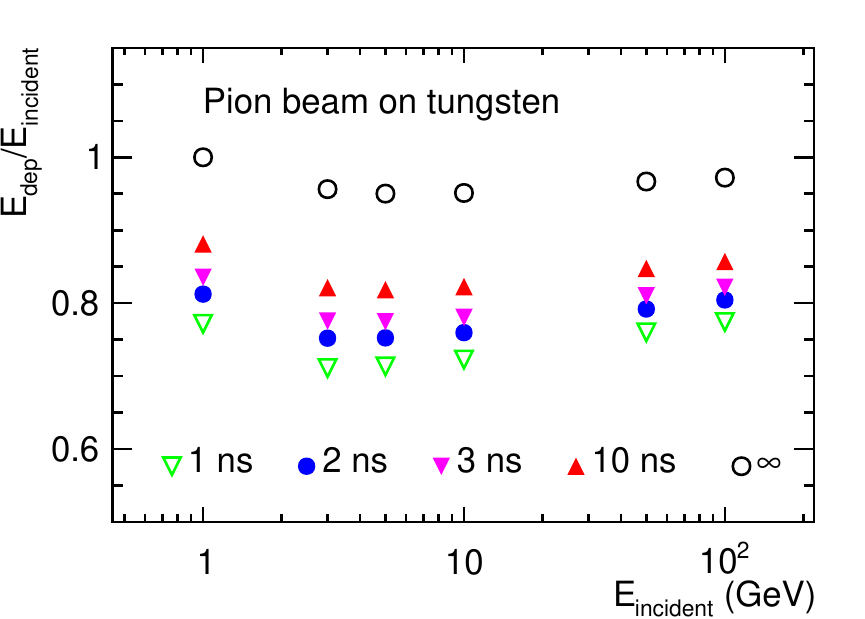}
    \caption{\label{fig:collection} Average fraction of deposited energy for various incident pion energies and integration time windows.}
\end{figure}

\begin{figure}
\centering
    \includegraphics[width=\columnwidth]{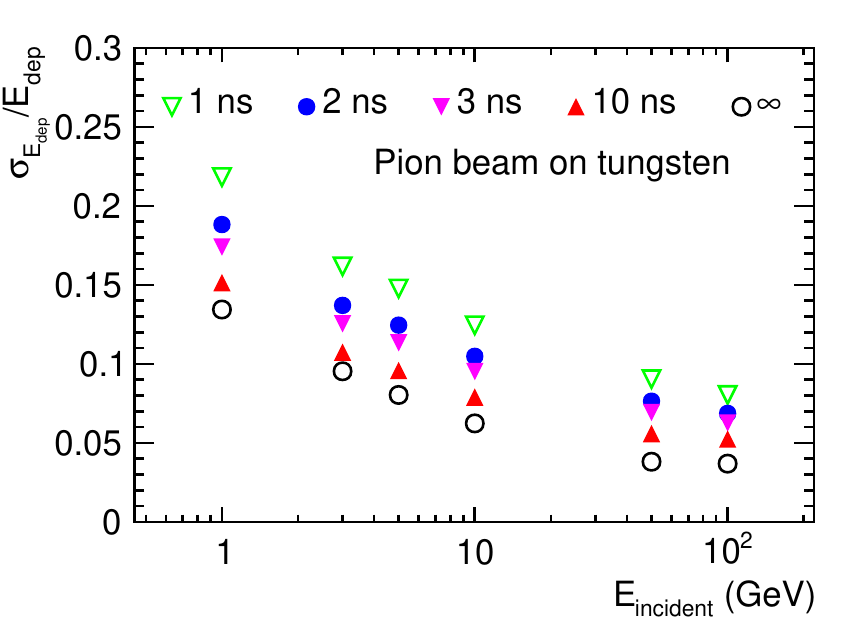}
    \caption{\label{fig:resolution} Relative RMS deviation of deposited energy for various incident pion energies and integration time windows.}
\end{figure}

The average deposited energy and RMS deviation from the average are 
calculated for different pion energies and integration time windows. Results of these 
studies are presented in Figs.\ \ref{fig:collection} and \ref{fig:resolution}. 
Even for an infinite integration time window a fraction of energy remains 
undetected due to escaping particles like neutrons and neutrinos, but also backscattered particles from the initial stages of the shower. Simulations are performed for two configurations of particle 
thresholds (see \secref{sec:mars}). First with the default thresholds and 
then with thresholds for the charged hadrons, muons, 
electrons, and photons ten times lower than the default and the neutron 
threshold at 1~meV. Energy deposition in tungsten is not changed with the 
modification of the thresholds for charged particles and gamma rays. 
However, when the neutrons are tracked down into the thermal energy range
the average deposited energy in tungsten increases by 8\% for infinite 
integration time. This increase is due to 
the contribution of $(n,\gamma)$ reactions at neutron energies below 0.1~MeV. 
Each such reaction in tungsten deposits about 6.5~MeV. This 
effect is not seen for integration times shorter than $\sim20\unit{ns}$. 

Between 70 and 80\% of the energy is deposited for $\tred < 1\unit{ns}$. 
Correspondingly, the ultimate calorimeter energy resolution is good even for very short integration times. Several real-detector effects, such as the sampling term or the fluctuation of detected signals, are neglected in this simulation. This simulation also assumes very fast readout of deposited energy. For a real calorimeter, the dependence of the resolution on the length of the integration time will be weaker, and short integration times, of the order of a few ns, are sufficient to reach energy resolution close to the optimal one.

\begin{figure}
\centering
    \includegraphics[width=.49\textwidth]{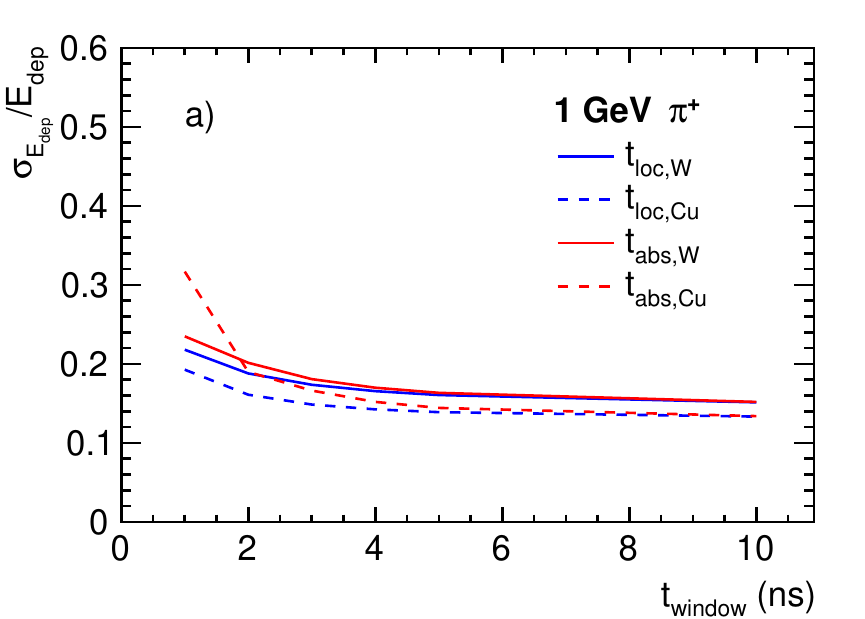}
    \includegraphics[width=.49\textwidth]{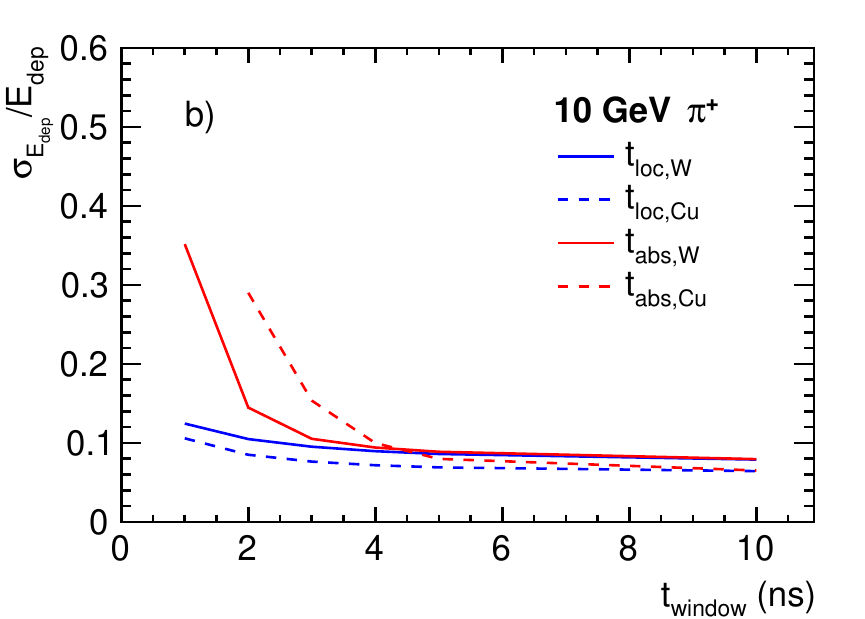}
    \includegraphics[width=.49\textwidth]{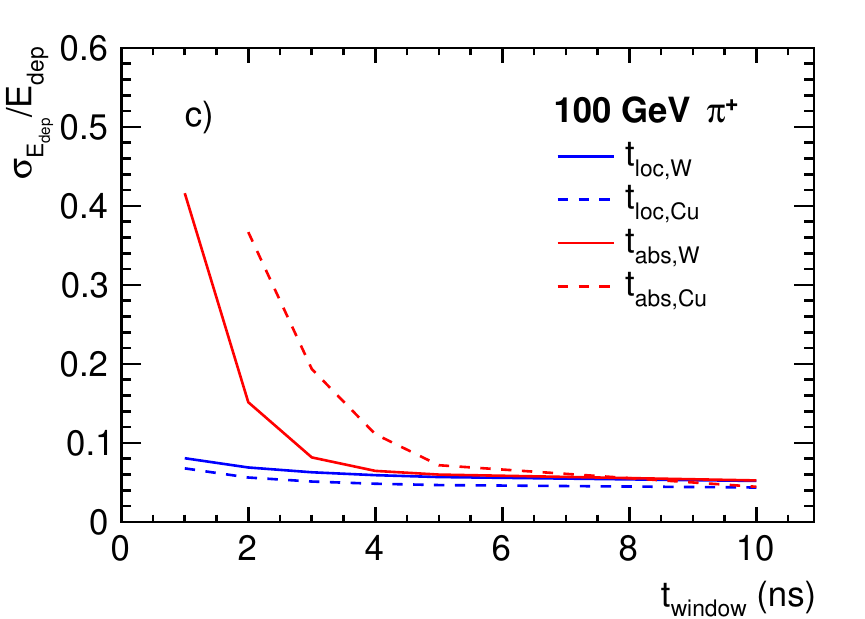}
    \caption{\label{fig:copper} Comparison of the energy resolution of copper 
    and tungsten “ideal” calorimeters for different pion energies and time windows:
    a) 1~GeV, b) 10~GeV, c) 100~GeV.}
\end{figure}

A calculation of the energy resolutions for an ``ideal'' copper calorimeter at different pion energies and integration times 
is also performed. The calorimeter has a cylindrical shape with 
length of 204~cm, corresponding to $\sim13\,\lint$ and a radius of 170~cm, corresponding to $\sim11\,\lint$. 
The results are presented in \figref{fig:copper} in 
comparison with resolutions of an ``ideal'' tungsten calorimeter. Resolutions in the case of integration in \emph{absolute} time \tabs, 
where the integration windows begin at the same time in all calorimeter layers, are also presented for comparison. 
It is clear that for short integration windows, the resolution obtained with integration in absolute time is significantly worse compared to the local time integration.

For a given absolute time window $\tabs$ the fraction of primary hadrons that produce only a MIP-like signal without starting a shower is $P = \exp(-\tabs c/\lint)$. 
As a result, for $\tabs$ of the order of 1-2~ns, the amplitude distribution contains a pronounced MIP-like component. 
Together with the fact that only a fraction of the shower energy is detected during a short time window, this leads to a deterioration of the energy resolution. 
This is not the case for \tred, where the beginning of the integration window depends on the thickness traversed by the shower.

\section{Conclusions}
  \label{sec:conclusions}
  Ultra-fast hadron calorimeters with integration time of a few ns can substantially reduce the effects of event pileup and other backgrounds in the particle physics experiments. 
Studies performed using a prototype of an iron-scintillator based calorimeter in 
the Fermilab test beam demonstrate that 80\% of the shower energy for pions and 
protons in the 30-120 GeV range is deposited within 2~ns. 
Test beam results of the hadronic shower properties are in good agreement with the MARS15 code simulation which is based on detailed propagation of the showers in matter. 
Using MARS15 simulations of the tungsten and copper calorimeters we demonstrate 
that a few ns energy integration time assures collection of over 80\% of the 
incident particle energy and the calorimeter energy resolution is similar to the 
case with infinite signal integration time. 
Fast detectors with response time of 1-2~ns sensitive to hadron shower energy deposits are required to utilize the unique potential of ultra-fast hadron calorimetry.

\section{Acknowledgments}
\label{sec:acknoledgments}

This document was prepared using the resources of the Fermi National Accelerator Laboratory (Fermilab), a U.S.\ Department of Energy, Office of Science, HEP User Facility. Fermilab is managed by Fermi Research Alliance, LLC (FRA), acting under Contract No. DE-AC02-07CH11359. 
The authors acknowledge the support received from the Ministry of Education, Science and Technological Development  (Republic of Serbia) within the projects OI171012 and OI171018. 
The authors wish to thank the staff of the Fermilab test beam facility for excellent support at all times during the experiment. 
We owe a special note of thanks to our late colleague Anatoly Ronzhin, who provided invaluable assistance with the setup for the test beam.

\section*{References}
\bibliography{bibliography/bibliography.bib}

\begin{thebibliography}{10}
\expandafter\ifx\csname url\endcsname\relax
  \def\url#1{\texttt{#1}}\fi
\expandafter\ifx\csname urlprefix\endcsname\relax\def\urlprefix{URL }\fi
\expandafter\ifx\csname href\endcsname\relax
  \def\href#1#2{#2} \def\path#1{#1}\fi

\bibitem{atlas_pII}
\mbox{ATLAS} Collaboration, \href{https://cds.cern.ch/record/2055248}{{ATLAS
  Phase-II Upgrade Scoping Document}}~(CERN-LHCC-2015-020, LHCC-G-166).
\newline\urlprefix\url{https://cds.cern.ch/record/2055248}

\bibitem{CMS_pI}
J.~Butler, D.~Contardo, M.~Klute, J.~Mans, L.~Silvestris,
  \href{https://cds.cern.ch/record/2055167}{{CMS Phase II Upgrade Scope
  Document}}~(CERN-LHCC-2015-019, LHCC-G-165), on behalf of the CMS
  collaboration.
\newline\urlprefix\url{https://cds.cern.ch/record/2055167}

\bibitem{CLIC_PhysDet_CDR}
L.~Linssen, A.~Miyamoto, M.~Stanitzki, H.~Weerts (Eds.), {{Physics and
  Detectors at CLIC: CLIC Conceptual Design Report}}, 2012, cERN-2012-003,
  ANL-HEP-TR-12-01, DESY-12-008, KEK-Report-2011-7.
\newblock \href {http://arxiv.org/abs/1202.5940} {\path{arXiv:1202.5940}},
  \href {http://dx.doi.org/10.5170/CERN-2012-003}
  {\path{doi:10.5170/CERN-2012-003}}.

\bibitem{Cald93}
A.~Caldwell, L.~Hervás, J.~Parsons, F.~Sciulli, W.~Sippach, L.~Wai,
  Measurement of the time development of particle showers in a uranium
  scintillator calorimeter, Nuclear Instruments and Methods in Physics Research
  Section A: Accelerators, Spectrometers, Detectors and Associated Equipment
  330~(3) (1993) 389 -- 404.

\bibitem{Vinz86}
M.~D. Vincenzi, et~al., Experimental study of uranium-scintillator and
  iron-scintillator calorimetry in the energy range 135–350 gev, Nuclear
  Instruments and Methods in Physics Research Section A: Accelerators,
  Spectrometers, Detectors and Associated Equipment 243~(2) (1986) 348 -- 360.

\bibitem{t3b}
C.~Adloff, et~al., The time structure of hadronic showers in highly granular
  calorimeters with tungsten and steel absorbers, Journal of Instrumentation
  9~(07) (2014) P07022.

\bibitem{Aco91}
D.~Acosta, et~al., Electron-pion discrimination with a scintillating fiber
  calorimeter, Nuclear Instruments and Methods in Physics Research Section A:
  Accelerators, Spectrometers, Detectors and Associated Equipment 302~(1)
  (1991) 36 -- 46.

\bibitem{Akc07}
N.~Akchurin, et~al., Measurement of the contribution of neutrons to hadron
  calorimeter signals, Nuclear Instruments and Methods in Physics Research
  Section A: Accelerators, Spectrometers, Detectors and Associated Equipment
  581~(3) (2007) 643 -- 650.

\bibitem{Mok95}
N.~V. Mokhov, The {MARS} code system User's Guide, Fermilab, 1995,
  {Fermilab-FN-628, \url{https://mars.fnal.gov}}.

\bibitem{Mok14}
N.~V. Mokhov, et~al., {MARS15} code development driven by the intensity
  frontier needs, Prog. Nucl. Sci. Technol. 4 (2014) 496--501.

\bibitem{ftbf}
\href{http://ftbf.fnal.gov/}{{Fermilab test beam facility}}.
\newline\urlprefix\url{http://ftbf.fnal.gov/}

\bibitem{Mash08}
S.~Mashnik, et~al., {CEM03.03 and LAQGSM03.03} event generators for the {MCNP6,
  MCNPX and MARS15} transport codes~(LANL Report LA-UR-08-2931),
  \arxivlink{0805.0751}.

\bibitem{Str05}
S.~I. Striganov, On the theory and simulation of multiple coulomb scattering of
  heavy charged particles, Rad. Prot. Dosimetry 116 (2005) 293--296.

\bibitem{Bicron}
\href{http://www.crystals.saint-gobain.com/sites/imdf.crystals.com/files/documents/sgc-bc400-404-408-412-416-data-sheet.pdf}{{Organic
  scintillation materials and assemblies}}, {Saint Gobain Crystals}.
\newline\urlprefix\url{http://www.crystals.saint-gobain.com/sites/imdf.crystals.com/files/documents/sgc-bc400-404-408-412-416-data-sheet.pdf}

\bibitem{Feu115M}
V.~Rykalin, et~al., Photo-multiplier with Improved Linearity on the Base of
  {FEU-115}, IHEP Protvino, 1995, preprint IHEP 95-12.

\bibitem{drs4}
\href{https://www.psi.ch/drs}{{DRS} chip webpage}.
\newline\urlprefix\url{https://www.psi.ch/drs}

\bibitem{ftbf-beam}
\href{http://ftbf.fnal.gov/beam-delivery-path/}{Beam structure at the
  {Fermilab} test beam facility}.
\newline\urlprefix\url{http://ftbf.fnal.gov/beam-delivery-path/}

\end{thebibliography}

\end{document}